\documentclass[twocolumn,amsmath,amssymb,aps,draft,showkeys,showpacs]{revtex4}

\usepackage{latexsym}

\begin{document}

\author{Sandro S. e Costa}\email{sancosta@astro.iag.usp.br}
\author{Reuven Opher}\email{opher@astro.iag.usp.br} 
\affiliation{Instituto de Astronomia, Geof\'\i sica e Ci\^encias
Atmosf\'ericas \\ Universidade de
S\~ao Paulo\\
R. do Mat\~ao, 1226 -- Cidade Universit\'aria\\
CEP 05508-900 -- S\~ao Paulo -- SP -- Brazil}
\title{Casimir energy in finite expanding universes: a prelude}
\date{December 6, 2002}

\begin{abstract}
We present the Green's functions that are the solutions of the massive
Klein-Gordon equation for a scalar field with non-minimal coupling to
gravitation for several static and expanding cosmological models. An
important feature of such functions is that they can be used to study the
appearence of a Casimir energy in models of the universe with compact
spatial sections.
\end{abstract}


\maketitle

\section{Introduction}

In the description of the universe based on General Relativity, the idea of
curved manifolds is essential. Einstein's equations are mathematical
relations between the curvature of spacetime and the matter and energy,
which act as a source of the curvature and can be seen as a first step in
the study of the structure of the physically possible models of the
universe. Given the curvature of a spacetime, it is important to
characterize its global properties, such as its topology. Since General
Relativity provides no clues about these properties, it is valid to include
in the list of possible models of the universe, those in which the space does
not have the simplest topology.

A space with non-trivial topology may be compact (i.e., finite),
regardless of its curvature. This means that the universe may present
certain peculiarities due to its finite size, such as periodic spatial
patterns. The observation of any of these peculiarities may indicate the
global topology of the universe, so that it is very important to know what
kind of mark a determined topology may imprint on the visible universe.
Cosmic topology, which deals with this issue, has been the subject of a
growing interest, reflected in the publication of recent reviews \cite
{Lachieze,Levin} and introductory texts \cite{Lachieze2,Levin2,Weeks}.

An interesting physical effect due to the finiteness of the space is the
appearance of a Casimir energy. The simplest case where this energy,
measurable experimentally, appears is between two parallel reflecting planes
in the Minkowski vacuum. However, this case can be generalized. The Casimir
energy can be seen to be created by the imposition of periodic boundary
conditions on a generic field propagating in the vacuum and can be studied
theoretically with a scalar field, by means of Green's functions $G\left(
x,x^{\prime }\right) $ for the Klein-Gordon equation \cite{BD}. The
procedure begins with the classical energy-momentum tensor associated with
the field, obtained from the Green's function through a second order
differential operator $D_{\mu \nu }^{x,x^{\prime }}$, 
\begin{equation}
\label{xis}T_{\mu \nu }\left( x,x^{\prime }\right) =D_{\mu \nu
}^{x,x^{\prime }}G\left( x,x^{\prime }\right) . 
\end{equation}
From this, we obtain the mean value of the field in the vacuum, such that 
\begin{equation}
\label{ipsilon}\left\langle 0\left| T_{\mu \nu }\right| 0\right\rangle =\lim
_{x\rightarrow x^{\prime }}T_{\mu \nu }\left( x,x^{\prime }\right) . 
\end{equation}
In general, when speaking of the Casimir energy, it is to the finite part of
the $00$ component of this mean value that one refers (see, e.g., \cite
{Daniel,Daniel2}).

As seen in some recent works \cite{Daniel,Daniel2}, the same kind of study
can be done in models of the universe with non-trivial topology. However,
these studies have been limited, until now, to static models of the
universe. Nevertheless, there is not, a priori, any reason not to include
models with expansion. What is important to notice is that a fundamental
step in this process is to write, for each kind of expansion of the
universe, the specific Green's functions for the Klein-Gordon equation.

The purpose of this work is to present the Green's functions found as
solutions of the generalized Klein-Gordon equation for a scalar field of
mass $m$ with non-minimal coupling $\xi $ to the gravitational field for
static and expanding cosmological models. In the expanding cases, the
functions presented are valid for a group of cosmological models,
parametrized by the curvature scalar $k$ and by an index $n$. Such a group
of models includes, for example, the de Sitter solution.

The Green's functions for universes with non-trivial topology are obtained
from the general solutions by noting that the Green's function for a generic
finite space, ${\cal M}=\widetilde{{\cal M}}/\Gamma $, is 
\begin{equation}
\label{intro}G_{{\cal M}}\left( x,x^{\prime }\right) =\sum_\Gamma G_{%
\widetilde{{\cal M}}}\left[ x,\Gamma \left( x^{\prime }\right) \right] , 
\end{equation}
where $\Gamma $ are the elements of the group which defines ${\cal M}$, each
one represented as an operator that identifies a point $P$ on the finite
space ${\cal M}$ and its image $P^{\prime }$ onto the infinite covering
space $\widetilde{{\cal M}}$, i.e., $P^{\prime }=\Gamma \left( P\right)
\equiv P$, with $P^{\prime }\in \widetilde{{\cal M}}$ and $P\in $ ${\cal M}.$
For 3-dimensional spaces of constant curvature the covering space $%
\widetilde{{\cal M}}$ can be \cite{Lachieze}
\begin{itemize}
\item the spherical space $\mathbb{S}^3$, of positive curvature; 
\item the flat space $\mathbb{E}^3$, of null curvature; 
\item or the hyperbolical space $\mathbb{H}^3$, of negative curvature.
\end{itemize}

The paper is organized as follows. The Klein-Gordon equation for a scalar
field is first written in a generalized form valid for the curved spacetimes
described in the Friedmann-Lema\^\i tre-Robertson-Walker line element of
standard cosmology. Some general Green's functions are then obtained for
specific cases of static and expanding spacetimes. Finally, some comments on
the results obtained are given. General solutions of the generalized
Klein-Gordon equation are presented in the appendix. Throughout this work,
natural units are used ($c=G=\hbar =1$).

\section{Initial considerations}

For a classical scalar field with a Lagrangian density 
\begin{equation}
\label{a}{\cal L}=\sqrt{-g}\left[ \frac 12\partial _\mu \varphi \partial
^\mu \varphi -\left( \frac 12\xi R\varphi ^2+\frac 12m^2\varphi ^2\right)
\right] , 
\end{equation}
one obtains the field equation 
\begin{equation}
\label{um}\Box \varphi +\left( \xi R+m^2\right) \varphi =0, 
\end{equation}
where $\Box \equiv \left[ -g\right] ^{-1/2}\partial _\mu \left( \sqrt{-g}%
g^{\mu \nu }\partial _\nu \right) $ is the d'Alembertian operator, $m^2$ and 
$\xi $ are constants, and $R=g_{\mu \nu }R^{\mu \nu }$ is the Ricci scalar
derived from the spacetime metric $g_{\mu \nu }$, with $g=\det \left[ g_{\mu
\nu }\right] $,

In this work the interest is focused on metrics representing cosmological
solutions, in particular, those with a Friedmann-Lema\^\i
tre-Robertson-Walker (FLRW) line element, 
\begin{equation}
\label{e}ds^2=dt^2-a^2\left( t\right) \left[ \frac{dr^2}{1-kr^2}+r^2\left(
d\theta ^2+\sin ^2\theta d\phi ^2\right) \right] , 
\end{equation}
which can be written as 
\begin{equation}
\label{f}ds^2=a^2\left( \eta \right) \left[ d\eta ^2-d\chi ^2-\frac{\sin ^2%
\sqrt{k}\chi }k\left( d\theta ^2+\sin ^2\theta d\phi ^2\right) \right] , 
\end{equation}
where $k=0,\pm 1$, with $t$ the cosmological time, $\eta $ the conformal
time, and $dt=ad\eta $. Assuming Einstein's equations with a cosmological
constant $\Lambda $, and matter a perfect fluid, this line element gives the
Friedmann equation 
\begin{equation}
\label{Friedmann}\frac{D^2}{a^2}\equiv \frac 1{a^2}\left( \frac 1a\frac{da}{%
d\eta }\right) ^2=\frac{8\pi }3\rho +\frac \Lambda 3-\frac k{a^2} 
\end{equation}
for the evolution of the scale factor $a$, with $\rho $ the energy density
of the matter, and 
\begin{equation}
\label{pressure}2\frac 1{a^3}\frac{d^2a}{d\eta ^2}+\frac k{a^2}=\Lambda
-8\pi p, 
\end{equation}
where $p$ is the pressure of the matter. For the FLRW metric, the Ricci
scalar is 
\begin{equation}
\label{g}R=\frac 6{a^2}\left[ \frac 1a\frac{d^2a}{d\eta ^2}+k\right] 
\end{equation}
and 
\begin{equation}
\label{l}\Box =\frac 1{a^2}\left[ \frac{\partial ^2}{\partial \eta ^2}%
+2D\frac \partial {\partial \eta }-\nabla ^2\right] . 
\end{equation}
Therefore, the field equation becomes 
\begin{equation}
\label{m}\frac 1{a^2}\left[ \frac{\partial ^2\varphi }{\partial \eta ^2}+2D%
\frac{\partial \varphi }{\partial \eta }-\nabla ^2\varphi +\left( \xi
R+m^2\right) a^2\varphi \right] =0. 
\end{equation}

\section{Green's functions}

A Green's function can always be constructed, using the eigenfunctions and
eigenvalues of the general solution of the homogeneous equation. However,
this is not always an easy task, especially in non-flat spacetimes, where,
for example, plane waves are replaced by generalized solutions to
Helmholtz equation \cite{KT} (see appendix). Therefore, since the Green
functions studied here must satisfy the more general equation, 
\begin{equation}
\label{mextra}\frac 1{a^2}\left[ \frac{\partial ^2G}{\partial \eta ^2}+2D%
\frac{\partial G}{\partial \eta }-\nabla ^2G\right] +\left( \xi R+m^2\right)
G=\frac{\delta \left( x-x^{\prime }\right) }{\sqrt{-g}},
\end{equation}
we search for solutions to Eq. (\ref{mextra}) in its homogeneous form,
keeping in mind that such solutions must be symmetric in $x$ and $x^{\prime }
$. In the appendix, some general solutions of the homogeneous field Eq.
(\ref{m}) are presented.

Assuming isotropy of the universe, the use of the coordinates $\left( \chi
,\theta ,\phi \right) $ allows for a choice of orientation where the two
points $x$ and $x^{\prime }$ are radially aligned, i.e., $\theta =\theta
^{\prime }$, $\phi =\phi ^{\prime }$, and $\chi ^{\prime }=0$, so that for
the FLRW line element, 
\begin{equation}
\label{nabla2}\nabla ^2=\frac{\partial ^2}{\partial \chi ^2}+2\sqrt{k}\cot 
\sqrt{k}\chi \frac \partial {\partial \chi },
\end{equation}
where $\chi $ is the spatial distance between the points $x$ and $x^{\prime }
$. Thus, the homogeneous equation to be solved is 
\begin{equation}
\label{Geq}\frac{\partial ^2G}{\partial \eta ^2}+2D\frac{\partial G}{%
\partial \eta }-\frac{\partial ^2G}{\partial \chi ^2}-\frac{2\sqrt{k}}{\tan 
\sqrt{k}\chi }\frac{\partial G}{\partial \chi }+\left( \xi R+m^2\right)
a^2G=0.
\end{equation}
Several different Green's functions obey this equation since there are
several boundary conditions that can be defined. For example, according to
different boundary conditions, we have the causal Green function (or Feynman
propagator) $G_F\left( x,x^{\prime }\right) $, the Wightman's functions $%
G^{+}\left( x,x^{\prime }\right) $ and $G^{-}\left( x,x^{\prime }\right) $
for positive and negative frequency, respectively, or the elementary Hadamard
function $G^{\left( 1\right) }\left( x,x^{\prime }\right) $. In quantum
field theory, these different functions are associated with mean values of
some field functions \cite{BD}.

\subsection{Static universes}

For flat spaces, the trivial solution of Friedmann's equation is Minkowski
space, which is a vacuum static solution for $\Lambda =0.$ In this simple
case, $a=1$, $R=0$, and 
\begin{equation}
\label{G0}\frac{\partial ^2G}{\partial \eta ^2}-\frac{\partial ^2G}{\partial
\chi ^2}-\frac 2\chi \frac{\partial G}{\partial \chi }+m^2G=0.
\end{equation}
Using the invariant 
\begin{equation}
\label{interval}s=\sqrt{\Delta \eta ^2-\chi ^2},
\end{equation}
where $\Delta \eta \equiv \eta -\eta ^{\prime }$, Eq. (\ref{G0}) becomes 
\begin{equation}
\label{Gs0}\frac{d^2G}{ds^2}+\frac 3s\frac{dG}{ds}+m^2G=0,
\end{equation}
with the general solution 
\begin{equation}
\label{GJN}G\left( x,x^{\prime }\right) =\frac 1s\left[ c_1H_1^{\left(
1\right) }\left( ms\right) +c_2H_1^{\left( 2\right) }\left( ms\right)
\right] ,
\end{equation}
where $H_\nu ^{\left( 1\right) }$ and $H_\nu ^{\left( 2\right) }$ are
Hankel's functions of order $\nu $ of the first and second kinds, respectively 
\cite{GR}. Another invariant, the world function $\sigma $ \cite
{Synge,Buchdahl,Roberts}, which in the Minkowski space is given as $\sigma
=s^2/2$, could have been used to give a similar general solution. Values for
the coeficients $c_1$ and $c_2$ are set in accordance with the particular
boundary conditions used in different Green's functions. For example, in the
case where $k=m=0$, Eq. (\ref{mextra}) is a wave equation with the massless
positive frequency Wightman function, 
\begin{equation}
\label{D0}D^{+}\left( x,x^{\prime }\right) =-\frac 1{4\pi ^2s^2}.
\end{equation}
Another type of Green's function, the massless Hadamard elementary function, 
\begin{equation}
\label{D1}D^{\left( 1\right) }\left( x,x^{\prime }\right) =-\frac 1{2\pi
^2s^2},
\end{equation}
is also a solution to Eq. (\ref{mextra}) for this case. Since near the
origin \cite{GR}, 
\begin{equation}
\label{Hs}H_1^{\left( 1\right) }\left( x\right) \simeq -\frac{2i}{\pi x}%
,\;\;H_1^{\left( 2\right) }\left( x\right) \simeq \frac{2i}{\pi x},
\end{equation}
we have 
\begin{equation}
\label{G0final}G^{+}\left( x,x^{\prime }\right) =\frac{im}{8\pi s}%
H_1^{\left( 2\right) }\left( ms\right) .
\end{equation}
This solution can be used as a limit that any other positive frequency
solution for curved spaces must reach.

For non-flat spaces ($k=\pm 1$), Friedmann's equation with a non-zero
cosmological constant has the vacuum ($\rho =0$) static solution,
\begin{equation}
\label{sextra3}a=\left( 3k\right) ^{1/2}\Lambda ^{-1/2},
\end{equation}
with $k\Lambda ^{-1}>0$. In this case, we have, from Eqs. (\ref{g}) and (\ref
{m}), 
\begin{equation}
\label{kGl}\frac{\partial ^2G}{\partial \eta ^2}-\frac{\partial ^2G}{%
\partial \chi ^2}-2\sqrt{k}\cot \sqrt{k}\chi \frac{\partial G}{\partial \chi 
}+\left( 6\xi +\frac{3m^2}\Lambda \right) kG=0,
\end{equation}
where $k=\pm 1$. If 
\begin{equation}
\label{GA}G\left( x,x^{\prime }\right) =\frac{\sqrt{k}\chi }{\sin \sqrt{k}%
\chi }A\left( x,x^{\prime }\right) ,
\end{equation}
Eq. (\ref{kGl}) becomes similar to (\ref{Gs0}), and thus 
\begin{equation}
\label{GB}G\left( x,x^{\prime }\right) =\frac{\sqrt{k}\chi }{\sin \sqrt{k}%
\chi }\frac 1s\left[ c_1H_1^{\left( 1\right) }\left( \mu s\right)
+c_2H_1^{\left( 2\right) }\left( \mu s\right) \right] ,
\end{equation}
and 
\begin{equation}
\label{GBessel}G^{+}\left( x,x^{\prime }\right) =\frac{i\mu }{8\pi }\frac{%
\sqrt{k}\chi }{\sin \sqrt{k}\chi }\frac 1sH_1^{\left( 2\right) }\left( \mu
s\right) ,
\end{equation}
where $s$ is given by Eq. (\ref{interval}), and $\mu ^2\equiv 3m^2\Lambda
^{-1}+\left( 6\xi -1\right) k$. Therefore, when $6\xi =1$, $m=0$, and $\chi
\simeq 0$, we again obtain Eq. (\ref{G0final}), as expected.

The spherical case, $k=1$, deserves special consideration since the universe
represented is finite spatially, so that there are several equivalent
spatial paths between the points $x$ and $x^{\prime }$, all of which must be
represented by the Green's function. Therefore, using the internal symmetry
of the space $\mathbb{S}^3$ in Eq. (\ref{GBessel}), we have 
\begin{equation}
\label{Gsphere}G^{+}\left( x,x^{\prime }\right) =\frac{i\mu }{8\pi }%
\sum_{n=-\infty }^\infty \frac{\chi _n}{\sin \chi }\frac{H_1^{\left(
2\right) }\left( \mu \sqrt{\Delta \eta ^2-\chi _n^2}\right) }{\sqrt{\Delta
\eta ^2-\chi _n^2}},
\end{equation}
where $\sin \chi _n\equiv \sin \left( \chi +2\pi n\right) =\sin \chi $ was
used. For any other spatially finite universe one uses an equivalent
procedure \cite{Daniel,Daniel2}.

\subsection{Expanding universes}

When $\Lambda =0$, a group of solutions of the Friedmann equation (\ref
{Friedmann}) can be found by means of the {\it ansatz} 
\begin{equation}
\label{h}\rho =Ca^{-n},
\end{equation}
where $C$ and $n$ are constants. This general group of solutions is then 
\begin{equation}
\label{adeeta}a\left( \eta \right) =\alpha ^{-\frac 2{2-n}}k^{\frac
1{2-n}}\csc ^{\frac 2{2-n}}\left[ \sqrt{k}\left( \frac n2-1\right) \eta
\right] ,
\end{equation}
where $\alpha ^2\equiv 8\pi C/3$ and $n\neq 2$. Thus we can write 
\begin{equation}
\label{de}D=\sqrt{k}\cot \left[ \sqrt{k}\left( \frac n2-1\right) \eta
\right] ,
\end{equation}
and 
\begin{equation}
\label{j}Ra^2=\left( 4-n\right) 3k\csc ^2\left[ \sqrt{k}\left( \frac
n2-1\right) \eta \right] .
\end{equation}
Using the {\it ansatz} given by Eq. (\ref{h}) we can solve Friedmann's
equation for $a$ and then substitute the expression found into Eq. (\ref
{pressure}), to obtain the linear equation of state $3p=\left( n-3\right)
\rho $.

For the case where $n=2$, the solution is 
\begin{equation}
\label{jextra1}a\left( \eta \right) =\exp \left[ \left( \alpha ^2-k\right)
^{1/2}\eta \right] ,
\end{equation}
with 
\begin{equation}
\label{jextra2}D=\left( \alpha ^2-k\right) ^{1/2};\;\;Ra^2=6\alpha ^2.
\end{equation}

Table \ref{tabela1} shows, for various integral values of the index $n$, the
expression for the scale factor $a$ as a function of the cosmological time $t$%
.

\begin{table}[t]
\begin{ruledtabular}
\begin{tabular}{ccc}
\multicolumn{1}{c}{$n$} & \multicolumn{1}{c}{$a\left( t\right) $} & 
\multicolumn{1}{c}{Comments} \\ \hline
$0$ & $\left( 2\alpha \right) ^{-1}\left[ e^{\alpha t}+ke^{-\alpha t}\right]  $ & exponential expansion\footnote{Exponential expansion is characteristic of the de Sitter's solution, a vacuum solution with a cosmological constant.} \\
$1$ & $2^{-2}{\alpha }^{2}t^{2} +k{\alpha }^{-2} $ & quadratic expansion \\ 
$2$ & $\sqrt{\alpha ^{2}-k}\;t $ & linear expansion\footnote{Linear expansion is characteristic of Milne's universe, where $k=-1$ and $\rho =0$.} \\
$3$ & $2^{-2/3}\left( 3\alpha \right) ^{2/3}t^{2/3} $ & matter\footnote{Only the solution with $k=0$, known as Einstein-de Sitter, is shown here. The more general solution can only be written as a function of the conformal time $\eta $.} \\ 
$4$ & $\sqrt{ 2\alpha t-kt^2} $ & radiation \\ 
\end{tabular}
\end{ruledtabular}
\caption{\label{tabela1} {Summary of some of the solutions
(not all physically relevant) of Friedmann's equation with
$\Lambda =0$ for the scale factor $a\left( t\right)$,
from the ansatz $\rho =3\alpha ^2\left( 8\pi\right)^{-1} a^{-n}$.}}
\end{table}

Although Eqs. (\ref{adeeta}) and (\ref{jextra1}) give a group of solutions
which are valid for any value of the parameter $n$, not all values of $n$
give easily solvable field equations. Here we consider only the cases where $%
n=0,2$ and $4$. For $n=2$ and $n=4$, only the massless ($m=0$) equations are
considered.

\subsubsection{Solution with $n=0$}

Since in this case $\rho $ is a constant, one has then a situation
completely equivalent to the case of a vacuum with the presence of a
cosmological constant $\Lambda =3\alpha ^2$. Therefore, the solution with $%
n=0$ is equivalent to the de Sitter solution, the curved spacetime most
studied by quantum field theorists \cite{BD}, and which is to cosmology
what a hydrogen atom is to atomic physics \cite{Gonzalez-Diaz} 
.

The de Sitter solution is, in fact, a group of solutions, each one with a
different $k$, but all related to portions of the 4-dimensional surface $%
\eta _{\mu \nu }x^\mu x^\nu =-\alpha ^{-2}$, which is an hyperboloid
embedded in a 5-dimensional Minkowski space of metric $\eta _{\mu \nu }=%
\mathrm{diag}\left[ 1,-1,-1,-1,-1\right] $. Different sections of this
surface represent the various de Sitter spacetimes, each one with a distinct
curvature. There also exists a static representation of de Sitter spacetime,
which will not be treated here \cite
{BD,Lord,Eriksen-Gron,Schrodinger,Hawking}. It is, however, worthwhile to
note that despite much investigation of the de Sitter spacetime, it is rare
to find descriptions of the hyperbolic de Sitter solution, which is not even
cited in some standard texts \cite{BD,Hawking}), perhaps due to an
`aesthetic' prejudice: ``the open model has the unnatractive feature that in
horospherical coordinates it leads to an uneasy distinction of the third
spatial component'' \cite{Grensing}.

For the flat case, we have 
\begin{equation}
\label{deSitter1}ds^2=\alpha ^{-2}\eta ^{-2}\left[ d\eta ^2-d\chi ^2-\chi
^2\left( d\theta ^2+\sin ^2\theta d\phi ^2\right) \right] ,
\end{equation}
with the corresponding parametrization 
\begin{equation}
\label{para1}\left\{ 
\begin{array}{l}
x^0=2^{-1}\eta -\left( \alpha ^{-2}+\chi ^2\right) \left( 2\eta \right)
^{-1} \\ 
x^1=-2^{-1}\eta -\left( \alpha ^{-2}-\chi ^2\right) \left( 2\eta \right)
^{-1} \\ 
x^2=-\left( \alpha \eta \right) ^{-1}\chi \sin \theta \cos \phi  \\ 
x^3=-\left( \alpha \eta \right) ^{-1}\chi \sin \theta \sin \phi  \\ 
x^4=-\left( \alpha \eta \right) ^{-1}\chi \cos \theta 
\end{array}
\right. \;\;\;,
\end{equation}
where $0\leq \phi \leq 2\pi $, $0\leq \theta \leq \pi $, $0\leq r<\infty $, $%
-\infty <\eta <0$. In this case, it is easy to see that the parametrization
of the hyperboloid is incomplete since $x^0+x^1\geq 0$.

In the spherical case, the line element is 
\begin{equation}
\label{deSitter2}ds^2=\alpha ^{-2}\csc ^2\eta \left[ d\eta ^2-d\chi ^2-\sin
^2\chi \left( d\theta ^2+\sin ^2\theta d\phi ^2\right) \right] ,
\end{equation}
with the relations 
\begin{equation}
\label{para2}\left\{ 
\begin{array}{l}
x^0=\alpha ^{-1}\cot \eta  \\ 
x^1=\alpha ^{-1}\csc \eta \cos \chi  \\ 
x^2=\alpha ^{-1}\csc \eta \sin \chi \sin \theta \cos \phi  \\ 
x^3=\alpha ^{-1}\csc \eta \sin \chi \sin \theta \sin \phi  \\ 
x^4=\alpha ^{-1}\csc \eta \sin \chi \cos \theta 
\end{array}
\right. \;\;\;,
\end{equation}
where $0\leq \phi \leq 2\pi $, $0\leq \theta \leq \pi $, $0\leq \chi \leq
\pi $, $0\leq \eta <\pi $.

Finally, for $k=-1$, we have 
\begin{equation}
\label{deSitter3}ds^2=\frac{\mathrm{csch}^2\,\eta }{\alpha ^2}\left[ d\eta
^2-d\chi ^2-\sinh ^2\chi \left( d\theta ^2+\sin ^2\theta d\phi ^2\right)
\right] \;,
\end{equation}
with 
\begin{equation}
\label{para3}\left\{ 
\begin{array}{l}
x^0=\alpha ^{-1}
\mathrm{csch}\,\eta \cosh \chi  \\ x^1=\alpha ^{-1}\coth \,\eta  \\ 
x^2=\alpha ^{-1}
\mathrm{csch}\,\eta \sinh \chi \sin \theta \cos \phi  \\ x^3=\alpha ^{-1}
\mathrm{csch}\,\eta \sinh \chi \sin \theta \sin \phi  \\ x^4=\alpha ^{-1}%
\mathrm{csch}\,\eta \sinh \chi \cos \theta 
\end{array}
\right. \;\;\;,
\end{equation}
where $0\leq \phi \leq 2\pi $, $0\leq \theta \leq \pi $, $0\leq \chi <\infty 
$, $-\infty <\eta <\infty $.

The invariant quantity to be considered in this case is not simply the
interval $s$, as defined in Eq. (\ref{interval}), but rather the geodesic
interval $\sigma \left( x,x^{\prime }\right) $ in the hyperboloid \cite
{Thurston,Coxeter}, a geometric quantity which is another example of a world
function \cite{Synge,Buchdahl,Roberts}, given in this case as \cite
{Tagirov2,Tagirov,Allen-Jacobson,Schomblond-Spindel} 
\begin{equation}
\label{worldfunction}\sigma \left( x,x^{\prime }\right) \equiv \alpha ^{-1}%
\mathrm{arccosh}\left| p\left( x,x^{\prime }\right) \right| ,
\end{equation}
where $p\left( x,x^{\prime }\right) \equiv -\alpha ^2\eta _{\mu \nu }x^\mu
\left( x^{\prime }\right) ^\nu $ is an auxiliary quantity. Using Eqs. (\ref
{para1}), (\ref{para2}), and (\ref{para3}) and the radial alignment between
the points $x$ and $x^{\prime }$, we can write 
\begin{equation}
\label{sigma}p\left( x,x^{\prime }\right) =\left\{ 
\begin{array}{l}
1+\left( \Delta \eta ^2-\chi ^2\right) \left( 2\eta \eta ^{\prime }\right)
^{-1} \\ 
1-\csc \eta \csc \eta ^{\prime }\left( \cos \Delta \eta -\cos \chi \right) 
\\ 
1+\mathrm{csch}\,\eta \,\mathrm{csch}\,\eta ^{\prime }\left( \cosh \Delta
\eta -\cosh \chi \right) 
\end{array}
\right. ,
\end{equation}
for $k=0$, $1$, and $-1$, respectively, so that, for any value of $k$, we
obtain 
\begin{equation}
\label{dp}\left( p^2-1\right) \frac{d^2G}{dp^2}+4p\frac{dG}{dp}+\left( 12\xi
+m^2\alpha ^{-2}\right) G=0,
\end{equation}
with the solution 
\begin{eqnarray}
\label{GdS}G\left( x,x^{\prime }\right) &=&c_1F\left( \frac 32-\nu ,\frac
32+\nu ;2;\frac{1-p}2\right)\\ \nonumber 
&+&c_2F\left( \frac 32-\nu ,\frac 32+\nu ;2;\frac{1+p}2\right) ,
\end{eqnarray}
where $\nu ^2\equiv 2^{-2}3^2-\left( 12\xi +m^2\alpha ^{-2}\right) $. An
analysis of the conditions that must be imposed on this solution to obtain
the values of $c_1$ and $c_2$ can be found in the literature \cite
{Schomblond-Spindel} and will not be reproduced here.

We note that Eq. (\ref{dp}) has also solutions in terms of the associated
Legendre functions \cite{Tagirov2,Tagirov}, 
\begin{equation}
\label{GdSL}G\left( x,x^{\prime }\right) =\left( p^2-1\right) ^{-1/2}\left[
c_1P_{-\frac 12+\nu }^1\left( p\right) +c_2Q_{-\frac 12+\nu }^1\left(
p\right) \right] ,
\end{equation}
or 
\begin{equation}
\label{GdPdp}G\left( x,x^{\prime }\right) =\frac d{dp}\left[ c_1^{\prime
}P_{-\frac 12+\nu }\left( p\right) +c_2^{\prime }Q_{-\frac 12+\nu }\left(
p\right) \right] ,
\end{equation}
valid since $p$ assumes only real values.

In the specific case of a massless field ($m=0$) with conformal coupling ($%
\xi =1/6$), we get 
\begin{equation}
\label{GdSD}D_{\xi =\frac 16}\left( x,x^{\prime }\right) =\frac{c_1p+c_2}{%
p^2-1},
\end{equation}
which gives 
\begin{equation}
\label{GdSD0}D_{\xi =\frac 16}^{+}\left( x,x^{\prime }\right) =-\frac{\alpha
^2}{8\pi ^2\left( p-1\right) }=-\left[ \frac{16\pi ^2}{\alpha ^2}\sinh ^2%
\frac{\alpha \sigma }2\right] ^{-1},
\end{equation}
a result formally similar to the one obtained for an uniformly accelerated
observer in Minkowski space (see also Eq. (\ref{D0})) \cite{BD}.

\subsubsection{Solution with $n=2$}

For $n=2$ we have the line element 
\begin{equation}
\label{Milne}ds^2=e^{2\eta \sqrt{\alpha ^2-k}}\left[ d\eta ^2-\frac{dr^2}{%
1-kr^2}-r^2\left( d\theta ^2+\sin ^2\theta d\phi ^2\right) \right] ,
\end{equation}
which, for $k=-1$ and $\alpha =0$, can be transformed in Minkowski space by
the change of coordinates 
\begin{equation}
\label{rt}\overline{r}=e^\eta r\,,\;\;t=e^\eta \sqrt{1+r^2}.
\end{equation}
This peculiar parametrization of the flat Minkowski space is known as the
Milne universe \cite{Narlikar,Rindler}.

In the general case where $n=2$, the homogeneous Eq. (\ref{Geq}) becomes
simpler in the flat massless case, i.e., with $k=m=0$, and we have 
\begin{equation}
\label{damped}\frac{\partial ^2G}{\partial \eta ^2}+2\alpha \frac{\partial G%
}{\partial \eta }-\frac{\partial ^2G}{\partial \chi ^2}-\frac 2\chi \frac{%
\partial G}{\partial \chi }+6\xi \alpha ^2G=0.
\end{equation}
This damped wave equation \cite{Stakgold}, has the solution 
\begin{equation}
\label{GG0}G\left( x,x^{\prime }\right) =e^{-\alpha \Delta \eta }G_0\left(
x,x^{\prime }\right) ,
\end{equation}
where $G_0\left( x,x^{\prime }\right) $ obeys the equation 
\begin{equation}
\label{G0eq}\frac{\partial ^2G_0}{\partial \eta ^2}-\frac{\partial ^2G_0}{%
\partial \chi ^2}-\frac 2\chi \frac{\partial G_0}{\partial \chi }+\left(
6\xi -2\right) \alpha ^2G_0=0,
\end{equation}
which is Eq. (\ref{G0}), for Minkowski space. Therefore, 
\begin{equation}
\label{Gdamped}G\left( x,x^{\prime }\right) =e^{-\alpha \Delta \eta }\frac
1s\left[ c_1H_1^{\left( 1\right) }\left( \beta s\right) +c_2H_1^{\left(
2\right) }\left( \beta s\right) \right] ,
\end{equation}
where $\beta ^2\equiv \left( 6\xi -2\right) \alpha ^2$ and $s$ is given by
Eq. (\ref{interval}). The non-flat massless cases can be obtained from this
solution in a similar way to that used for static non-flat cases.

\subsubsection{Solution with $n=4$}

Although for $n=4$, we have $Ra^2=0$ from Eq. (\ref{j}), it is not easy to
find a Green's function in this case. However, it is interesting to see that
in the homogeneous massless ($m=0$) equation there exists a symmetry between
the coordinates $\eta $ and $\chi $, i.e., 
\begin{equation}
\label{Gn4}\frac{\partial ^2G}{\partial \eta ^2}+2\sqrt{k}\cot \sqrt{k}\eta 
\frac{\partial G}{\partial \eta }=\frac{\partial ^2G}{\partial \chi ^2}+2%
\sqrt{k}\cot \sqrt{k}\chi \frac{\partial G}{\partial \chi }.
\end{equation}

\section{Final remarks}

We presented the Green's functions for a number of interesting cosmological
solutions. Not all plausible possibilities for scalar fields have been
covered and there remains a number of physically possible cosmological
solutions, for example, the expanding solutions with $n=2$ or $n=4$, for
which there is not yet a complete, easy available, analytical treatment.
Even where many studies have been made, there are lapses, such as the almost
complete absence of references to the hyperbolic de Sitter solution.
It is the intention of this article to stimulate more work in
this field.

In the context of cosmic topology, the task of writing Green's functions is
a preparatory step, and can be done analytically. The bulk of the work, the
use of these functions in cosmic topology, is mainly numerical and, for this
reason, is not presented here. 

The importance of studies linked to the field
of cosmic topology must not be underestimated. Despite some recent claims
about the beginning of a new era of `precision cosmology', we are still
unable to determine whether the universe is finite or not. Moreover, with
the observations being made currently, we cannot tell for sure whether the
universe has a really null curvature or just a very small one, which would be
a remnant from a pre-inflation era. In this context, finding topological
properties of the universe is very important. In the theoretical search for
these properties, the study of a scalar cosmological Casimir effect in
finite spaces, while not providing the final answer about the shape of the
universe, would certainly supply important information.

\section*{Acknowledgements}

S.S. e Costa thanks the Brazilian agency FAPESP (Funda\c c\~ao de Amparo \`a
Pesquisa do Estado de S\~ao Paulo) for financial support (grant 00/13762-6).
Both S.S. e Costa and R. Opher thank FAPESP (grant 00/06770-2) and the
Brazilian project PRONEX/FINEP (41.96.0908.00) for partial support. R. Opher
also thanks CNPq for partial support.

\appendix*

\section{Solutions for the field equation}

A Green's function can always be written using the eigenfunctions and
eigenvalues of the homogeneous equation it obeys \cite{Stakgold,Arfken}.
However, to obtain Green`s functions by this process for the finite spaces
of non-trivial topology is not an easy task, since the eigenvalues in
general are not easy to find. Therefore, it is more convenient to write a
Green`s function for infinite space and force it to satisfy the symmetries
of the finite space numerically. Nevertheless, even the functions for the
infinite space are not easy to obtain in this fashion because the
eigenfunctions are, in general, very complicated. In this appendix,
eigenfunctions for some specific cosmological solutions are presented.

The field equation (\ref{m}) has general solutions of the kind 
\begin{equation}
\label{n}\varphi \left( \eta ,x\right) =u\left( \eta \right) v\left(
x\right) ,
\end{equation}
where $u$ and $v$ are functions satisfying the equations 
\begin{equation}
\label{eta0}\frac{d^2u}{d\eta ^2}+2D\frac{du}{d\eta }+\left[ \ell ^2+\left(
\xi R+m^2\right) a^2\right] u=0,
\end{equation}
and 
\begin{equation}
\label{p}\nabla ^2v=-\ell ^2v,
\end{equation}
the Helmholtz equation.

\subsection{Spatial modes}

To obtain information about the size and shape of the universe, we must
study the spatial modes of the general solutions of the field equation. A
general form for the solutions of the Helmholtz equation, without taking
into account the finiteness of the universe, is \cite{BD} 
\begin{equation}
v\left( x\right) =\left\{ 
\begin{array}{lll}
\exp \eta _{ij}\ell ^ix^j\,, & \ell ^2=\eta _{ij}\ell ^i\ell ^j & \left(
k=0\right)  \\ 
\Pi _{\ell J}^{\left( \pm \right) }\left( \chi \right) Y_J^M\left( \theta
,\phi \right) , & \ell =\left( \ell ,J,M\right)  & \left( k=\pm 1\right) 
\end{array}
\right. \,,
\end{equation}
where $Y_J^M$ are the usual spherical harmonics (with limitations on the
possible values of $\ell ,J,M$), and 
\begin{equation}
\Pi _{\ell J}^{\left( -\right) }\left( \chi \right) =c_{\ell J}\sinh ^J\chi
\left( \frac d{d\cosh \chi }\right) ^{1+J}\cos \ell \chi ,
\end{equation}
with $\Pi _{\ell J}^{\left( +\right) }\left( \chi \right) $ obtained by
replacing $\ell $ with $-i\ell $ and $\chi $ with $-i\chi $ \cite{BD}.

We note that the solutions given above are not the only possible options:
for 3-D flat spaces, there are 11 coordinate systems in which the Helmholtz
equation is separable \cite{Eisenhart,Arfken}, while for hyperbolic spaces,
there are at least 12 such systems \cite{Costa1}. Such diversity is
particularly important in universes with compact spatial sections. In these
universes the coordinate system ($\chi ,\theta ,\phi $) is not always a good
choice since the characteristic symmetries of the spatial sections of such
universes generally do not fit in these coordinates, especially in the
hyperbolic case. An example where this occurs is in the calculation of the
volumes and other related functions of the polyhedra that serve as the
fundamental cells of compact hyperbolic spaces and the use of the
coordinate set ($\chi ,\theta ,\phi $) makes the problem almost intractable 
\cite{eCosta2}.

\subsection{Non-spatial modes}

Trying a solution of the kind 
\begin{equation}
\label{q}u\left( \eta \right) =a^pU\left( \eta \right) ,
\end{equation}
where $p$ is a constant, in Eq. (\ref{eta0}), we obtain 
\begin{eqnarray}
\label{r}\frac{d^2U}{d\eta ^2}+2\left( p+1\right) D\frac{dU}{d\eta }+\left[
\ell ^2+\left( \xi R+m^2\right) a^2\right] U \nonumber \\ 
+p\left[ \left( p+1\right)D^2+\frac 1a\frac{d^2a}{d\eta ^2}\right] U=0. 
\end{eqnarray}

In the case where $p=-1$, we obtain 
, using Eq. (\ref{g}), 
\begin{equation}
\label{sextra}\frac{d^2U}{d\eta ^2}+\left\{ \ell ^2+k+\left[ \left( \xi
-\frac 16\right) R+m^2\right] a^2\right\} U=0.
\end{equation}
This equation, which clearly becomes simpler for the case of a conformal
coupling, $\xi =1/6$, can be solved analytically for several cases of
cosmological interest.

\subsubsection{Static universes}

For Minkowski space, $a=1$, $k=R=0$ and 
\begin{equation}
\label{UMink}U=c_1\sin \left[ \left( \ell ^2+m^2\right) ^{1/2}\eta \right]
+c_2\cos \left[ \left( \ell ^2+m^2\right) ^{1/2}\eta \right] .
\end{equation}

For $k\neq 0$, we have from Eqs. (\ref{g}) and (\ref{sextra}), 
\begin{equation}
\label{sextra4}\frac{d^2U}{d\eta ^2}=-\left( \ell ^2+6\xi +m^2\frac{3k}%
\Lambda \right) U\equiv -\lambda ^2U,
\end{equation}
with 
\begin{equation}
U=c_1\sin \lambda \eta +c_2\cos \lambda \eta \,.
\end{equation}


\subsubsection{Expanding universes}

When $\Lambda =0$ and the scale factor $a$ is given by Eq. (\ref{adeeta}),
the function $u$ satisfies the equation 
\begin{equation}
\label{eta1}\frac{d^2u}{d\eta ^2}+2D\frac{du}{d\eta }+\frac{3k\xi \left(
4-n\right) u}{\sin ^2\left[ \sqrt{k}\left( \frac n2-1\right) \eta \right] }%
+\left( \ell ^2+m^2a^2\right) u=0.
\end{equation}
Then, 
\begin{widetext}
\begin{equation}
\label{modos}\frac{d^2U_{k,n}}{d\eta ^2}+
\left\{ k+\ell ^2+\frac{k\left( 4-n\right)
\left( 6\xi -1\right) }{2\sin ^2\left[ \sqrt{k}\left( \frac n2-1\right) \eta
\right] }+\frac{m^2 \alpha ^{-\frac 4{2-n}}k^{\frac 2{2-n}}}{\sin
^{\frac 4{2-n}}\left[ \sqrt{k}\left( \frac n2-1\right) \eta \right] }%
\right\} U_{k,n}=0, 
\end{equation}
\end{widetext}
where the expression for $a$ given in Eq. (\ref{adeeta}) was used.

\subsubsection*{Solutions for $n=0$}

When $n=0$, Eq. (\ref{modos}) becomes 
\begin{equation}
\label{uk0}\frac{d^2U_{k,0}}{d\eta ^2}+\left[ k+\ell ^2+\frac{\left( \frac
14-\nu ^2\right) k}{\sin ^2\sqrt{k}\eta }\right] U_{k,0}=0,
\end{equation}
where $\nu ^2\equiv 2^{-2}3^2-\left( 12\xi +m^2\alpha ^{-2}\right) $. The
solutions to this equation are 
\begin{equation}
\label{v}U_{0,0}=\sqrt{\eta }\left[ c_1J_\nu \left( \ell \eta \right)
+c_2N_\nu \left( \ell \eta \right) \right] ,
\end{equation}
for $k=0$ and 
\begin{eqnarray}
\label{d1}U_{1,0}=\sin ^{\frac 12}\eta \left[ c_1P_{-\frac 12+\sqrt{\ell ^2+1}}^\nu \left( -\cos \eta \right)\right. \nonumber\\
+\left. c_2Q_{-\frac 12+\sqrt{\ell ^2+1}}^\nu
\left( -\cos \eta \right) \right] 
\end{eqnarray}
and 
\begin{eqnarray}
\label{uh0}U_{-1,0}=\sinh ^{\frac 12}\eta \left[ c_1P_{-\frac 12+\sqrt{\ell
^2-1}}^\nu \left( \cosh \eta \right)\right.\nonumber\\
+\left. c_2Q_{-\frac 12+\sqrt{\ell ^2-1}}^\nu
\left( \cosh \eta \right) \right] 
\end{eqnarray}
for $k=1$ and $k=-1$, respectively.

\subsubsection*{Solutions for $n=2$}

When $n=2$, Eq. (\ref{sextra}) is 
\begin{equation}
\label{duk2}\frac{d^2U_{k,2}}{d\eta ^2}+\left[ \ell ^2+k+m^2e^{2\eta \sqrt{%
\alpha ^2-k}}+\left( 6\xi -1\right) \alpha ^2\right] U_{k,2}=0,
\end{equation}
which has the solution 
\begin{equation}
\label{uk2}U_{k,2}=c_1J_\gamma \left[ \frac{me^{\left( \alpha ^2-k\right)
^{1/2}\eta }}{\left( \alpha ^2-k\right) ^{1/2}}\right] +c_2N_\gamma \left[ 
\frac{me^{\left( \alpha ^2-k\right) ^{1/2}\eta }}{\left( \alpha ^2-k\right)
^{1/2}}\right] ,
\end{equation}
where 
\begin{equation}
\label{gamma}\gamma ^2\equiv -\left( \alpha ^2-k\right) ^{-1}\left[ \ell
^2+k+\left( 6\xi -1\right) \alpha ^2\right] .
\end{equation}
This solution reduces to 
\begin{equation}
\label{uk2m0}U_{k,2}^{m=0}=c_1\sin \left( \gamma \eta \sqrt{\alpha ^2-k}%
\right) +c_2\cos \left( \gamma \eta \sqrt{\alpha ^2-k}\right) 
\end{equation}
when $m=0$.

\subsubsection*{Solutions for $n=4$}

When $n=4$, Eq. (\ref{modos}) becomes 
\begin{equation}
\label{n4}\frac{d^2U_{k,4}}{d\eta ^2}+\left[ k+\ell ^2+\frac{m^2\alpha ^2}%
k\sin ^2\sqrt{k}\eta \right] U_{k,4}=0,
\end{equation}
whose solution for flat space is 
\begin{eqnarray}
\label{y}U_{0,4}=\frac 1{\sqrt{\eta }}\left[ c_1M_{\ell ^2\left( 4im\alpha
\right) ^{-1},4^{-1}\sqrt{7}}\left( im\alpha \eta ^2\right)\right.\nonumber\\
+\left. c_2M_{\ell
^2\left( 4im\alpha \right) ^{-1},4^{-1}\sqrt{7}}\left( im\alpha \eta
^2\right) \right] , 
\end{eqnarray}
where 
\begin{equation}
\label{z}M_{\lambda ,\mu }\left( z\right) =z^{\mu +1/2}e^{-z/2}\Phi \left(
\mu -\lambda +\frac 12,2\mu +1;z\right) 
\end{equation}
is the Whittaker $M$ function \cite{GR}. This solution can also be written
using the parabolic cylinder function $D_p\left( z\right) $ \cite{GR}.

For $k=1$ and $k=-1$, Eq. (\ref{n4}) yields a Mathieu equation, 
\begin{equation}
\label{g1}\frac{d^2U_{1,4}}{d\eta ^2}+\left[ \ell ^2+1+\frac{m^2\alpha ^2}2-%
\frac{m^2\alpha ^2}2\cos 2\eta \right] U_{1,4}=0,
\end{equation}
and a modified Mathieu equation, 
\begin{equation}
\label{k1}\frac{d^2U_{-1,4}}{d\eta ^2}+\left[ \ell ^2-1-\frac{m^2\alpha ^2}2+%
\frac{m^2\alpha ^2}2\cosh 2\eta \right] U_{-1,4}=0,
\end{equation}
respectively.

In the massless case, Eq. (\ref{n4}) has a single solution for all $k$'s, 
\begin{equation}
\label{uk4m0}U_{k,4}^{m=0}=c_1\sin \left[ \left( k+\ell ^2\right) ^{1/2}\eta
\right] +c_2\cos \left[ \left( k+\ell ^2\right) ^{1/2}\eta \right] .
\end{equation}

\end{document}